# Amateur Drone Monitoring: State-of-the-Art Architectures, Key Enabling Technologies, and Future Research Directions

Zeeshan Kaleem and Mubashir Husain Rehmani, *Senior Member, IEEE*

**Abstract**

The unmanned air-vehicle (UAV) or mini-drones equipped with sensors are becoming increasingly popular for various commercial, industrial, and public-safety applications. However, drones with uncontrolled deployment poses challenges for highly security-sensitive areas such as President house, nuclear plants, and commercial areas because they can be used unlawfully. In this article, to cope with security-sensitive challenges, we propose point-to-point and flying ad-hoc network (FANET) architectures to assist the efficient deployment of monitoring drones (MDr). To capture amateur drone (ADr), MDr must have the capability to efficiently and timely detect, track, jam, and hunt the ADr. We discuss the capabilities of the existing detection, tracking, localization, and routing schemes and also present the limitations in these schemes as further research challenges. Moreover, the future challenges related to co-channel interference, channel model design, and cooperative schemes are discussed. Our findings indicate that MDr deployment is necessary for caring of ADr, and intensive research and development is required to fill the gaps in the existing technologies.

## I. INTRODUCTION

THE development of the mini-drones, officially called unmanned air vehicles (UAVs) have captured the attention of hobbyists and the investors [1]. Drones have endless commercial applications, due to their relatively small size and ability to fly without an on-board pilot such as in agriculture, photography, surveillance, and numerous public services as shown in Fig.1.

With all these applications of drones one question becomes obvious: How we know that these drones are safe? Moreover, the number of media reports about incidents involving UAVs also increases with the increase in drone usage. For example, they have been used to bring smuggled goods into prisons, a drone hit power lines in Hollywood cutting off electricity for 600-700 customers, and White House reported security threats when a DJI Phantom II quadcopter crashed in its grounds. So, how to make sure these drones will not enter the No-Fly zone? and how to avoid collisions among drones? The other problem is what would happen if terrorists attempted to use amateur drone (ADr) for evil purposes? Moreover, if ADr drone is found in the No-Fly zone, then how can it be detected, tracked, localized, jammed, and hunted to stop its harmful consequence on the security-sensitive areas? In order to cope with these security threats, monitoring drones (MDr) deployment is required for surveillance, hunting, and jamming of the ADr. The main motivation of deploying MDr is to keep eye on the ADr which can lead to serious disasters in case no precautionary measures is taken on time. The most important aspect of MDr deployment is related to the MDr architecture because it should be self-configured in case of emergency situation without the help of the central ground control station (GCS).

In this article, we propose MDr architectures for different security situations. For less security-sensitive areas point-to-point architecture is proposed which has only one MDr to detect,track, and jam the ADr. Moreover, for highly security-sensitive areas, the flying ad-hoc network (FANET) architecture is proposed to intelligently estimate the situation and perform self-healing actions.

## II. MOTIVATIONS AND PROBLEMS IN THE EXISTING MDR SCHEMES

The increasingly usage of MDr poses challenges like robust and efficient detection, tracking, intruder localization, and jamming of ADr. The accuracy of detection is a basic requirement of the system. In general, the accurate detection is time-consuming. In fact, a precise moving object detection method makes tracking more reliable and faster, and supports correct classification, which is quite important for MDr detection to be successful.

**Industrial aspects**: To fulfill the demands discussed for amateur drone surveillance, industrialists are developing efficient detection, tracking, localization, and jamming solutions. For example, Dedrone has developed a cost effective technical solution by developing a DroneTracker and counter drone system. DroneTracker effectively protects the airspace by providing early-warning detection and alert after detecting drone in the space. It consists of arrays of sensors, high definition cameras, analysis

Z. Kaleem and M. H. Rehmani is with the Department of Electrical Engineering, COMSATS Institute of Information Technology, Wah, Pakistan e-mail: zeeshankaleem@gmail.com,mshrehmani@gmail.com

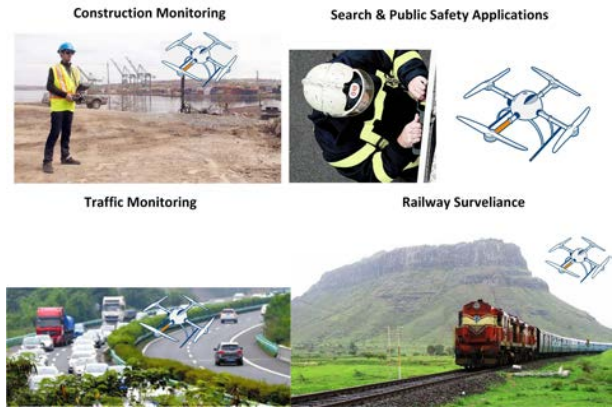

Fig. 1. Drones Applications: Construction monitoring, public-safety, traffic and railway monitoring.

and intelligent pattern recognition to detect and classified the drones. Similarly, Teleradio is a global technology leader in developing advanced anti-terrorist security systems. It recently developed SkyDroner product which has the capability to detect, distract and disable any ADr from flying into a security-sensitive areas. Hence, numerous ADr surveillance solutions exists in the industry but to improve their efficiency and accuracy more research and development is desired.

**Research and development aspects**: Detection techniques can be mainly classified by features, appearance, or motion. Based on these characteristics, firstly here we discuss the motion detection scheme. In literature [2], motion detection schemes are categorized as: 1) background subtraction; 2) spatio-temporal filtering; and 3) optical flow. Among these methods, background subtraction algorithms are most popular, because they are relatively simple in computing a static scene. However, the background is assumed to be static in this method. Thus, shaking cameras, waving trees, lighting changes are quite probable to cause serious problems to a background subtraction model. In addition, a successful background subtraction method needs to model the background as accurate as possible, and to adapt quickly to the changes in the background. These requirements add extra complexity to the computation of the model. Spatio-temporal filtering consider the motion pattern of the moving object throughout the frame sequence; but is highly sensitive to noise and variations in the movement pattern. The optical-flow based approach is robust to motion of camera and the ADr movement because in this approach relative movements between the observer and the scene is considered. The optical-flow transforms one image into the next image in a sequence of images. Moreover, it tells that how images in a sequence change to create the next image. It is therefore necessary to have two subsequent images to calculate optical-flow. Hence, in [3] optical-flow is suggested as benchmark analysis scheme for ADr motion detection using the MDr camera.

The existing motion detection algorithms have problem of high computational cost and less robustness because of changing extrinsic and intrinsic camera parameters. These changing parameters includes pan, tilt, translation, rotation, and zooming. Hence, most accurate and cost-effective motion detection algorithms are desired to take care of these parameters. To fill this gap, the adaptive algorithms by using the hybrid approaches needs to be proposed for efficient motion detection. So, in general there must be two main parts of the motion detection algorithm: firstly the concept of artificial flow should be introduced based on the movement of the camera, and secondly it must be compared with the real-optical flow to cover the image discrepancies. The choice of using two optical flows is motivated by the need to enhance the differences produced by dynamic objects. Moreover, the machine learning algorithms using the characteristics of the electromagnetic waves, sound, images are also desired to efficiently detect the ADr.

The next major step after detecting the ADr is localization and tracking of the ADr and intruder. To accurately estimate the position of the ADr and its intruder, the 3D position estimation algorithms are desired to be developed for more accurately position estimation of the ADr. The existing works utilized the Kalman filtering approach for accuarately determined the position of the drones in case of global positioning service (GPS) fail to work [4]. The importance of using Kalman filtering is highlighted in these works. Moreover, the utilization of the commercial frequency bands for MDr also possess the challenges of interference management with the existing system.

The last important step after detection, localization, and tracking is the jamming and hunting of the ADr. The detailed steps which includes the detection, localization, tracking, jamming, and hunting of ADr is depicted in Fig. 2. The details of jamming and hunting is also discussed later in this article.

## III. Potential Architectures and Deployment Scenarios for Monitoring Drones (MDr)

The suitable architecture for deployed MDr is necessary for monitoring the ADr in the No-Fly zones. The MDr architecture should be dynamic and can be created in an ad-hoc manner anywhere and anytime with or without the help of any centralize infrastructure. The distinction feature of the proposed architecture is that its functional capabilities are tailored for the context of meeting different security-situations, as well as efficient utilizing the available spectrum. This architecture have the potential

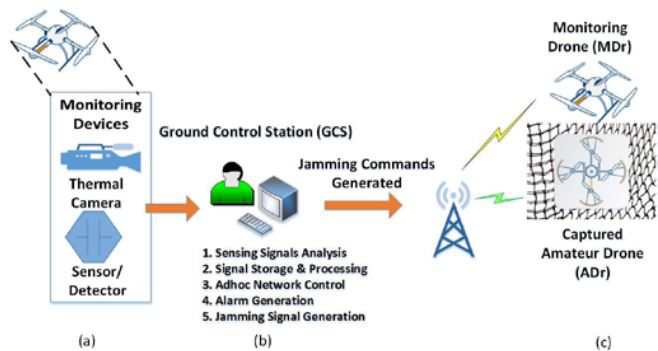

Fig. 2. Amateur drone capturing system step-wise representation: (a) data collection by using MDr; (b) ground control station for detection and jamming signal generation; (c) jamming and capturing of IDr.

to work in centralized and distributed manner. Based on this classifications, we divided the MDr architecture into four main kinds: 1) point-to-point (P2P) architecture with only ground control station (GCS) for line-of-sight (LOS) communication (GCS-P2P), 2) GCS with cellular or satellite stations assisted network architecture (GCS-C/SA), 3) GCS with mesh network based infrastructure (GCS-MA), and 4) GCS with software-defined networking based architecture (GCS-SDN). The important entities of these architectures are introduced below:

**Ground control station (GCS)**: MDr can operate with different levels of controls, but usually they are remotely controlled from a GCS. The GCS works as a central managing point of activity during MDr missions and provides necessary capability to plan and execute security-sensitive missions. The GCS has functions, such as mission planning, image processing, modules for mobile and satellite communications management, and displays to monitor the progress. The GCS encapsulates high speed processors for signal storage, processing, and analysis of data received from MDr. It also generates the alarm and jamming commands based on the security alerts.

**Monitoring drones (MDr) payload**: MDr carries payloads in terms of various types of sensors, thermal cameras, and other control information processing units. These sensors are operated and controlled from the ground. The data obtained from it is utilized for accurate detection, tracking, and jamming of ADr.

**Internet backhaul/Wireless backhaul**: The wired internet link such as Ethernet or fiber optics can be utilized to carry the data received from wireless radio access network (RAN) part to the GCS as shown in Fig. 3. This wired link can be replaced by wireless link by using high frequency which would have the capability to carry more data and also can be more reliable because of less interference as it will operate in different frequency band as the RAN link. So, the concept of flexible backhaul can be used which is one of the future research directions.

Moreover, we discuss four architectures mentioned above with their applications and suitability for different monitoring situations.

### A. GCS-P2P: Point-to-Point (P2P) Architecture with Ground Control Station (GCS)

The P2P architecture is proposed for MDr which only has the LOS communication link between the MDr and the GCS. This architecture can be deployed for the monitoring of less security-sensitive areas. For example, P2P-GCS architecture is the simplest architecture and is suitable for situation with MDr located in LOS position and near to the GCS, and only one MDr would be enough to detect, control, and track the ADr because of less security sensitivity of the area. In this scenario, connectivity is maintained over dedicated links to MDr, so control signal is more reliable with low latency. But unfortunately, the GCS-P2P architecture is not suited for dynamically varying environments and non-line-of-sight (NLOS) communication. It is not suitable for NLOS communication because P2P architecture is designed for low-range LOS communications. For NLOS communication the walls and bad weather conditions can block the signal to MDr due to high penetration loss. Hence, to use this architecture for NLOS communication at longer ranges, the GCS requires high-power transmitter and steerable antennas. Therefore, this will be inefficient solution in terms of power consumption and will also require costly hardware. The other limiting factors for GCS-P2P architecture is the significantly higher bandwidth requirements to support more number of MDr in the vicinity of GCS because each MDr will require independent LOS channel for connectivity. Moreover, the communication among MDr will always be from GCS similar as star network topology that will not be able to exploit the benefits of cooperation among MDr.

### B. GCS-C/SA: GCS with Cellular or Satellite Stations assisted Network Architecture

The GCS-P2P architecture discussed previously has the problem of smaller coverage and cannot work well for monitoring of large coverage area. For GCS-C/SA architecture shown in Fig.3 (b), the cellular/satellite provides better coverage for MDr as compared to GCS-P2P becuuause MDr can handover to other GCS which also enable the possibility of NLOS communication. The GCS-C/SA based MDr network can more efficiently connect the MDrs in the vicinity by taking the benefits of longer coverage and licensed spectrum. However, this connectivity can result in severe co-channel interference among the MDr and the cellular/satellite users using the same bandwidth. So, it will result in relatively poor data delivery using cellular/satellite network. This problem can be solved by deploying next-generation (5G) ultra-dense cellular base stations [5], but will result in increase of the system cost.

### C. GCS-MA: GCS with Mesh Network based Architecture

The MDr can utilize the GCS with mesh network based architecture (GCS-MA) to solve the challenges raised in GCS-C/SA architecture. In GCS-MA architecture as shown in Fig. 3 (c), each node (i.e. MDr) can act as a relay to forward data. The communication between MDr and a GCS can take place over several hops through intermediate MDr. This will result in shorter communication range which simplifies the link requirements and bandwidth can be reused more frequently and thus more efficiently. This will result in more efficient communication among MDr because of possibility of direct link as well as using mesh network based communications link. In result, the ADr located among the MDr can be easily detected, tracked, localized, and jammed with more accuracy. However, GCS-MA always requires intermediate MDr nodes to serve as a relay to communicate from GCS. Hence, to maintain such a link, efficient adaptive routing algorithms are desired which can adopt themselves based on the quickly changing environment of the MDr.

From these architectures, we conclude that for security-sensitive areas we need to deploy GCS-MA architecture for MDr because this can support the ad-hoc network of MDr which in turn have the benefits of low-cost, high coverage, low-latency, scalability, and high reliability as compared with GCS-P2P and GCS-C/SA architectures.

### D. GCS-SDN: Software-Defined Networking based Ground Control Station Architecture for MDr

The GCS for the three architecures discussed above has distributed control on all MDr where each MDr controls its behavior as shown in Fig. 4 (a). The deployment of MDr in the security-sensitive area will be more beneficial if it can be controlled and managed centrally rather than managing independently. This deployment would facilitate flexible deployment and management of new services and will help to have the overall view of the network situation. Previously, deployment of software-defined network (SDN) was focused on infrastructure-based static deployment scenario because it was believed that SDN is suitable only for static networks. But the recent research results [6] proved that the SDN deployment is beneficial for dynamically changing environment with moving/flying base stations. Thus, based on these results we proposed to deploy the SDN-based ground control station (GCS-SDN) assisted MDr for dynamically varying wireless environments as shown in Fig. 4 (b).

The SDN-based architecture partitioned the control and data plane of network with all the control functionalities moved to the cloud and are running as the applications on the top of SDN controller in the GCS-SDN. This can be easily managed centrally and updated easily from anywhere with more efficiency. Moreover, this will allow the operators to centrally control and decide the travelling path of MDr based on deployment situation. The GCS-SDN can also work efficiently in conjunction with the cellular networks because a centralized SDN controller can enable efficient radio resource and mobility management [7], which is important to efficiently utilize the available frequency band by avoiding interference. Furthermore, the GCS-SDN can be beneficial for public-safety scenarios where the malfunctioning base stations traffics can be by-passed by properly working base station because of having complete network information in the cloud. Moreover, in future the existing MDr architecture discussed previously shown in Fig. 3 can also be upgraded to SDN-based architectures for more efficient operations of MDr FANET.

## IV. Promising Key Technologies for ADr Monitoring

### A. Operational Frequency Bands for MDr Deployment

According to the latest reports by Federal Aviation Administration (FAA) [8], there is a forecast for year 2020 that 7 million drones will fly in the United States. This will result in intense competition of spectrum usage because drones are classified into two types based on their applications: 1) MDr and 2) ADr. Thus, the frequency band selection for MDr is more important for its secure operation from the jammer attacks and to have less interference from the nearby drones. Moreover, there are two main types of communication links: one is for control commands and the other is for payload/data communications between MDr and GCS. The control commands usually requires low-data rate communications links whereas for payload communications high data rate is desired to transfer the information like videos and sensor data. Currently, no specific frequency allocations have been made on an international level for control and payload communications. Thus, the precise selection of operational frequency band for MDr deployment is too important that can be different according to MDr deployment situation. Usually, for less security-sensitive areas MDr can operate on freely available industrial, scientific and medical (ISM) frequency bands of 2.4 or 5.8 GHz and remote control bands (433MHz, 800MHz) [9] for both control and payload/data communications. Since, ISM band is used for wireless networks so it sometimes results in loss of control over the MDr in densely populated areas because of interference with other wireless signals.

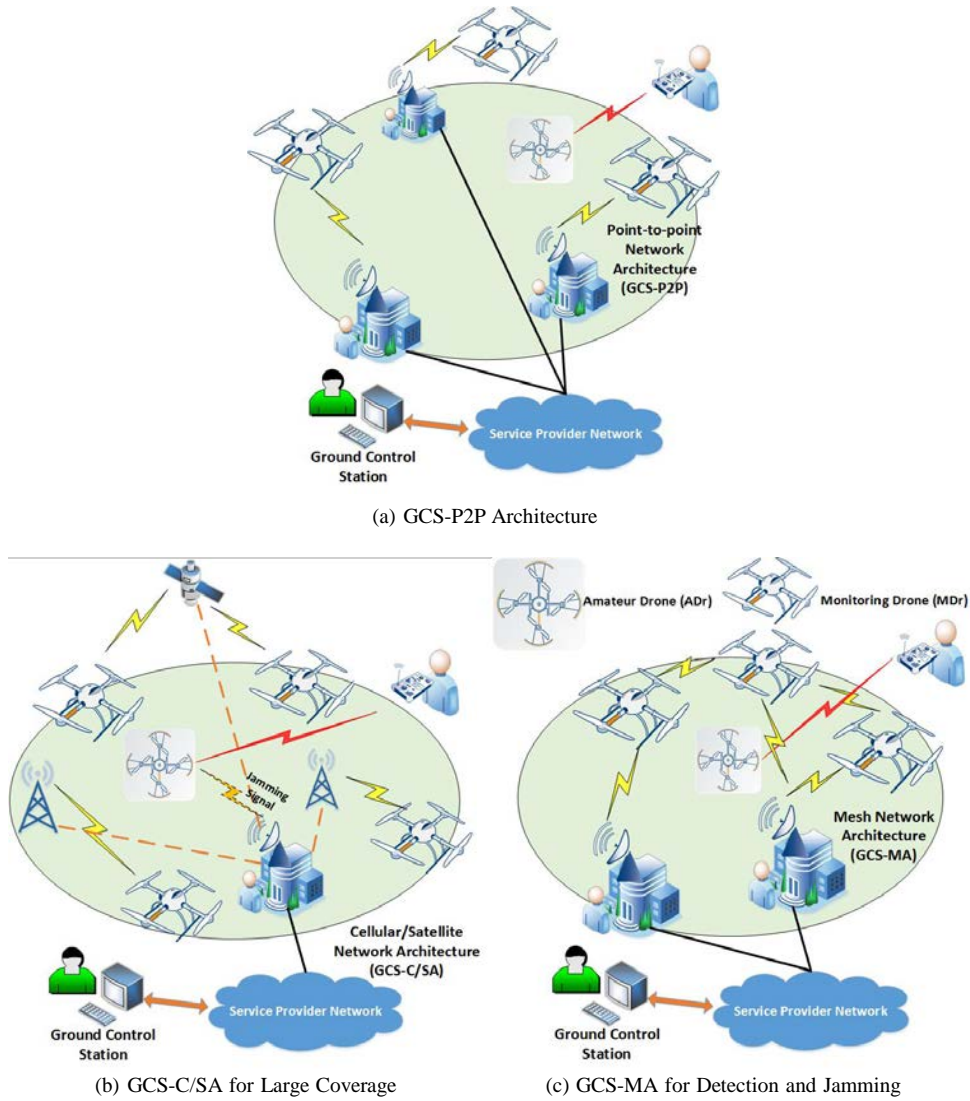

(a) GCS-P2P Architecture

(b) GCS-C/SA for Large Coverage

(c) GCS-MA for Detection and Jamming

Fig. 3. Monitoring drone architecture for controlling amateur drones flying in security-sensitive areas: (a) GCS-P2P architecture; (b) GCS-C/SA for large coverage; (c) GCS-MA for efficient detection and jamming.

For highly sensitive areas, MDr control and payload communication is recommended to use use the dedicated licsensed bands such as IEEE L-band and IEEE S-band [10] to avoid interference among other wireless communications systems. Additionally if governmental organizations want to use satellites to control their drones far from their own territories. In this case, MDr global positioning services (GPS) bands (GPS L1 1575.42MHz, GPS L2 1227.60MHz) [9] can be used for MDr control and payload operation. The pros and cons of utilizing different frequency bands for MDr are summarized in Table I. Moreover, we conclude that there is yet no dedicated spectrum available for MDr deployment in case of security sensitive areas, and thus based on the deployment situation it can be adopted. Hence, to internationally accommodate the usage of MDr for surveillance of ADr, significant efforts and research is desired to properly allocate the spectrum for MDr. Moreover, the frequency band could be decided in upcoming (2019) world radio communication conferences (WRC), organized by international telecommunication union (ITU) after every four year.

### B. Detection Technologies for ADr Monitoring

Drones with numerous applications and its possible security threats are discussed above in details. Hence, on-time detection of ADr in No-fly zone or national security-sensitive area is the first and most important step of security. The detection schemes can be broadly divided into three major groups: 1) electromagnetic waves-based detection, 2) sound waves detection for the drones which do not emit any radio/electromagnetic waves, and 3) drone detection using thermal imaging cameras. These techniques can have the ability to detect small drones in low-flying zone and non-cooperative targets in high clutter environments. The efficiency and range of detection schemes can be increased if the sound sensing, imaging methods, and radio waves detection are properly combined. The conventional RADAR cannot be used for ADr detection because of it usually designed for large aircraft. Moreover, mostly the ADr are made from plastic material which have dielectric property close to air, and thus have little or no reflection back to the transmitter. Hence, the efficient algorithms are desired for the proposed architecture discussed above for ADr detection.

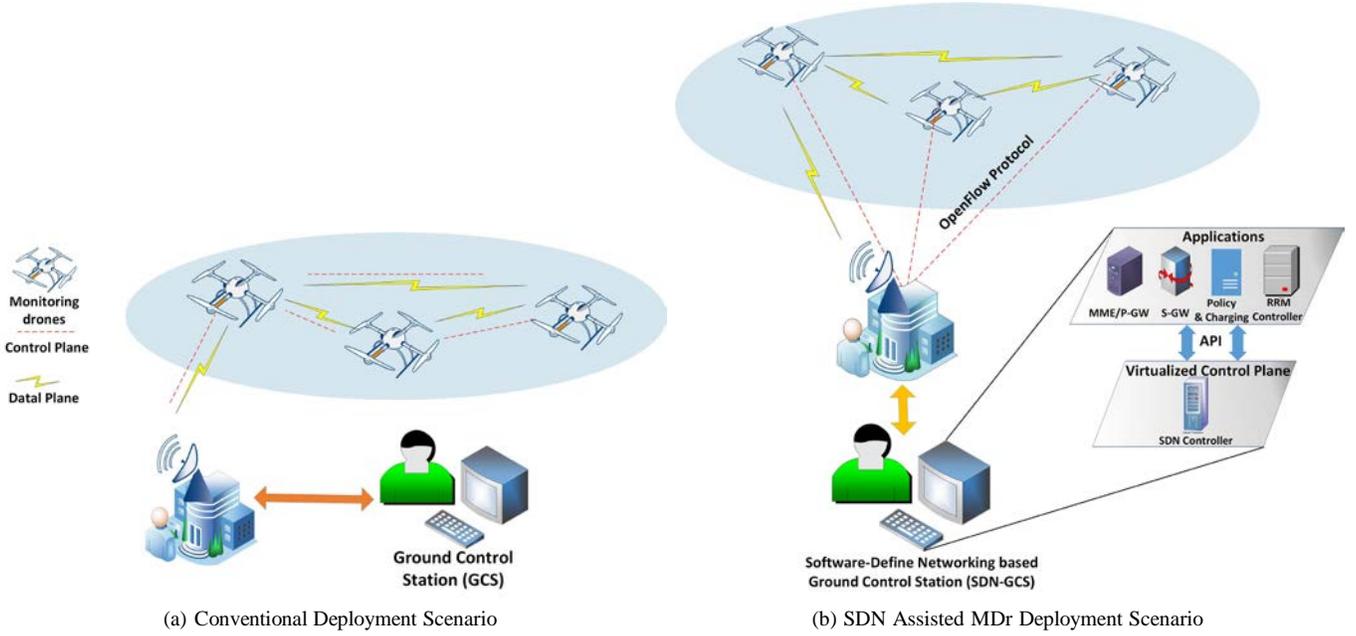

(a) Conventional Deployment Scenario         (b) SDN Assisted MDr Deployment Scenario

Fig. 4. MDr deployment scenarios: (a) Conventional deployment scenario; (b) SDN assisted MDr deployment scenario.

TABLE I
PROS AND CONS OF USING DIFFERENT FREQUENCY BANDS FOR MONITORING DRONE

| Frequency Band | Pros | Cons |
|---|---|---|
| 433MHz [9] | • Multiple channels available at one location for control and data<br>• Little multipath distortion<br>• Range is less susceptible to weather conditions<br>• Long transmission distance | • Does not support broadcast of quality video<br>• Legally transmitted power<br>inside band 433,050- 434,79 MHz<br>restricted to 10mW<br>effective radiated power (ERP) |
| 2.4GHz [9] | • Enables video broadcast<br>• Low cost system | • Expected interference in populated areas<br>due to the wide use: microwave<br>ovens, cordless phones, wireless LAN<br>• LOS operational requirements<br>• Range affected by humidity in the air<br>• Legal transmitting power restrictions<br>(100mW ERP) for the 802.11.b/g technology |
| 5.8GHz [9] | • Enables video broadcast<br>• Small transmitting antennas<br>• Low multipath distortions in spread spectrum OFDM modulation | • Severe multipath distortion causes<br>• Very poor performance in FM mode<br>• LOS operational requirements<br>• Range affected by humidity in the air |
| 1575.42MHz (GPS L1)<br>1227.60MHz (GPS L2)[9] | • Less interference with existing<br>Wireless communications system<br>• High coverage • Available free of charge<br>without any subscription or license | • Ionosphere and troposphere leads to<br>slow down of signal propagation speed |
| Any Bands using<br>Cognitive Radios Technology [10] | •Dynamic spectrum access<br>• Suitable for public-safety<br>and emergency situations<br>• Opportunistic spectrum utilization based<br>on commercial applications requirements | • To handle the tradeoff between<br>sensing and transmission |

**Linear Predictive Coding-based Sound Detection Scheme**: Drones have distinct sound as compared to the sound of car, thunder, and other flying or moving objects. Thus, based on these differentiating factors sound can be used for ADr detection. The main challenge in application of the sound detection schemes for ADr detection is the noisy and complex environment because the ADr sound could not be easily separable from the noise. The existing sound detection scheme used the linear predictive coding (LPC) scheme for sound detection. The LPC scheme is frequently used for speech recognition and works on the principle of detecting spikes in the frequency spectrum. The drones have spikes as well and thus it was chosen as the base for the sound detection. The LPC is used such that the sound at a point can be approximated in time from past samples.

These approximated coefficients are called LPC coefficients, and these coefficients can be trained for various speeds in time and stored in data base. But still there is a chance that similar sounds like drones could be falsely detected. To solve this problem, authors in [11] also considered the slope of the frequency spectrum to decrease the number of false alarm and they proved it was beneficial it terms of false alarm reduction.

**Sound Feature Detection and Classifications Algorithms**: The sound detection scheme works in two main phases, first there is feature extraction and then feature classifications by using machine learning algorithms. For sound feature extraction, the algorithms such as harmonic line association (HLA), mel-frequency cepstral coefficient (MFCC), and wavelet-based feature extraction are frequently used in audio signal processing. HLA technique arranges spectrum peaks, with level greater than noise level, into families of harmonically related lines that exceed the noise level into families of harmonically related narrow band lines. Thus, the HLA algorithm is also suitable for ADr sound detection because ADr also emit strong harmonic lines produced by propeller, and these harmonic lines will prove the presence of ADr. Similarly, MFCC has frequently used in speech recognition systems, and increasingly finding uses in applications such as sound similarity measures which can be useful for feature detection.

To detect and classify different sounds based on their features, the short-time Fourier transform (STFT) is used to obtain the spectrogram of the drone, car, and thunder sounds stored in our database. The variations in spectrum of frequencies can be seen by spectrogram that gives a visual representation of the frequencies. By using these frequencies we can extract features of sounds and then this can help to detect the drones. The spectrogram of three different type of sounds such as drone sound, car sound, and thunder sound is plotted in Fig. 5. The horizontal axis of spectrogram represents time, the vertical axis is frequency, and the intensity bar represents the amplitude of a particular frequency in time.

These spectrograms can be analyzed through HLA feature extraction algorithm by extracting a set of feature vectors and comparing them against the acoustic target database to find the label of the object (that is, either it is ADr sound, car sound, or thunder sound). For classifying the extracted feature from the spectrogram, we use machine learning (ML) algorithms. In ML, different set of algorithms are programmed in GCS that analyze the data obtained from MDr and try to make predictions about it. These algorithms are classified based on how learning is performed such as: 1) supervised learning and 2) unsupervised learning. In supervised learning, the algorithms are given a set of input and output data and then a model is developed based on input-output relationship. Then, a new set of input data from MDr is gathered and fed into the learned model so that the algorithm can make its predictions. The reason for selecting supervised learning based algorithms is that it includes the trained data set that consists of both features and labels irrespective of unsupervised learning models which only have features. Thus, by using these trained data sets, we have a task to use the set of features to predict the label of an object. The supervised learning based ML algorithms includes linear classifiers, support vector machine (SVM), and hidden markov model (HMM). In SVM, the situation is modeled by creating a finite-dimensional vector space, where each dimension represents a feature of a particular object such as bird or ADr. HMM is the statistical Markov model where the states (bird or ADr) are hidden. Each state can emit an output (sound), which is observed. By using the Bayes theorem we can obtain the probability that whether it was ADr or bird. The detailed steps of feature extraction till the classification is depicted in Fig. 5.

From the above discussions and operators recommendations, we conclude that the sound detection technologies range is not a fixed value but usually lies in between 50m-300m. The exact value of sound detection technologies depends on the parameters such as selected operational frequency, noise volume, and on the background noise of the drone.

**HGH Thermal Camera-based ADr Detection Technology**: The infrared (IR) thermal imaging has already been proved quite effective for industrial applications. In order to detect small flying drones IR thermal imaging technology could also be quite much effective. An infrared imaging camera works on a principle to detect the infrared heat energy and converts it into an electronic signal, then this signal is processed to produce a thermal image. Recently, HGH Infrared Systems has developed number of IR thermal cameras such as SPYNEL-X, SPYNEL-S, SPYNEL-C, and SPYNEL-U. These IR thermal cameras are equipped by sensors with different sensitivity level and camera that can take the panoromic images of various qualities. Hence, the IR thermal cameras MDr detection range suggested by different operator varies from 50m-400m based on these factors. They also have the capability to discriminate the target over extremely large areas even in darkness, fog or smoke. The SPYNEL 360-degree camera head takes a panoramic image of an entire wide area and built- in advanced algorithms can automatically detect and track an unlimited number of air, land, and maritime targets including large thermal engine UAVs. To further improve the performance, the research on advanced algorithms are desired which can efficiently detect these targets.

### C. ADr Localization Technologies

The main step after detecting the ADr is to localize and track it. The localization stands for finding the real-world position of the ADr in real-time. The accurate localizations of ADr is needed for its tracking, hunting, and future processing. For localization, the MDr and GCS needs the exact information about ADr altitude, ADr GPS coordinates, and the intruder GPS coordinates. To accurately locate ADr and position of intruder using low-quality inertial measurement unit (IMU) and by using global positioning system (GPS) in bad weather would be a challenging tasks as GPS accuracy decreases in bad weather. Although, some robust inertial/GPS localization are proposed in [12], but these schemes still have less precision and nearby signal interference are not modeled in this proposal. In order to solve this problem, we propose the solution to increase the

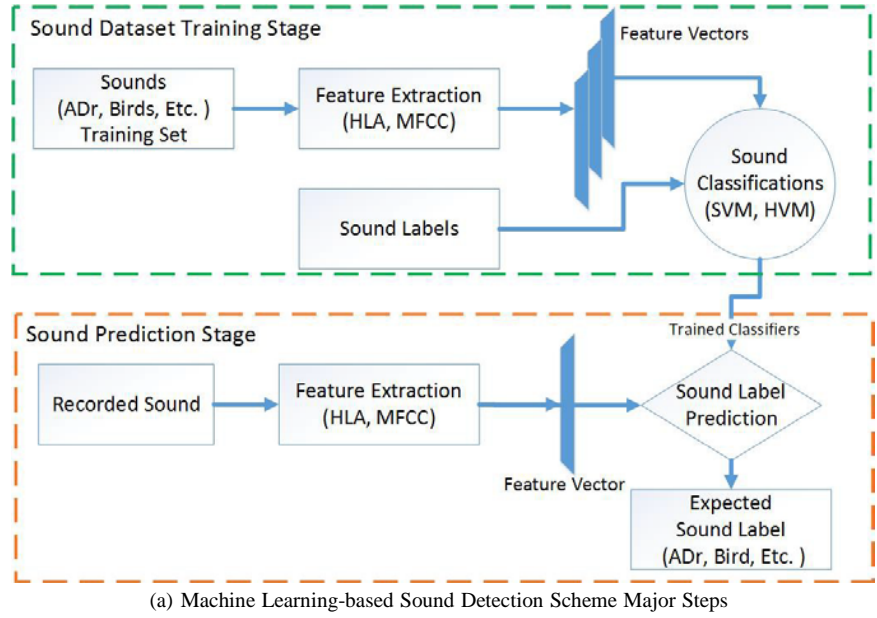

(a) Machine Learning-based Sound Detection Scheme Major Steps

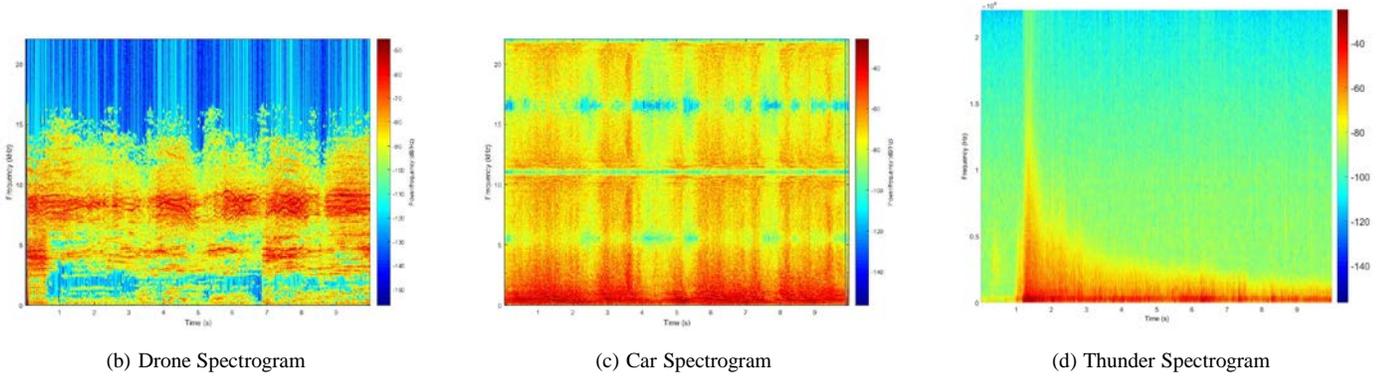

(b) Drone Spectrogram        (c) Car Spectrogram        (d) Thunder Spectrogram

Fig. 5. Machine learning-based sound detection scheme and spectrograms for sound classifications: (a) Sound detection using machine learning; (b) drone spectrogram; (c) car spectrogram; (d) thunder spectrogram.

accuracy of localization by creating the FANET of MDr deployed for monitoring purpose. The measured information from all the MDr would be used to more accurately locate the position of the ADr. Moreover, in order to increase the accuracy of the localization scheme, the images taken from the installed camera on MDr can be used beside GPS to obtain more accurate localization results. But this poses more research challenges like image drift effect and are prone to errors due to wrong estimation. They are also sensitive to illumination changes and environmental modifications. Hence, further research needed to solve these localization challenges and to increase the reliability of the existing schemes.

### D. ADr Tracking Technologies

Target tracking by using MDr is an important research area after detecting ADrs, especially the target tracking in 3-D space. MDr tracking capabilities depends on many aspects but we focus on the how accurately MDr gets the ADr information using the installed sensors on MDr and then how accurate and effective control commands are generated by ADr to achieve tracking. The first important step in ADr target tracking is to get full channel state information of the ADr but there is a chance that this information can be affected by external disturbances such as wind and other factors. The channel state information desired for ADr tracking includes the location, range, azimuth and elevation angles of the ADr. Numerous classical guidance and control algorithms caught researchers attentions in recent years such as vector field approach, Serret-Frenet representation based methods, line of sight algorithms, feedback linearization, and the back-stepping approach. But these approaches has the problem that they demand full and accurate channel state information at the MDr. Since, we do not have full state information available, so to efficiently utilize these tracking algorithms the estimation algorithms such as Kalman filter approach and extended Kalman filter algorithms are desired because these can accurately track the target without having full channel state information. Hence, more research is needed to design Kalman filter estimation based control algorithms which can efficiently work with limited channel state information and can track a dynamic 3D moving targets. Moreover, they have the capability to cope with the environmental changes such as the measurements corrupted by white noise and by stationary colored noise.

*E. MDr Routing Technologies*

FANET is also an ad-hoc network for UAVs but their requirements are relatively different from the traditional networking model, such as mobile ad-hoc networks (MANET) and vehicular ad-hoc networks (VANET). Hence, the FANET has different routing challenges based on the specific scenario, network situation and performance criteria as compared with the MANET and VANET. As we are focusing on routing protocols for MDr which should also be secure in order to save it from the intruder attacks. Previously, the FANET communications protocols that includes physical (PHY), medium access (MAC), and network layers are discussed [13] but here the survey of different routing protocols are provided without considering the MDr situations and security issues. So, based on these lacking we presented some challenges for future research. The main difference between the MDr FANET and the existing ad-hoc networks is the mobility. The speed of MDr is in between 30-460 km/h, thus this leads to several routing challenges like frequent update in MDr network topology because if one MDr is in outage or have battery constraints.

The notable point here is that except the routing technologies, as routing is only possible in GCS-MA architecture, the rest of the technologies such as detection, localization, and tracking are architecture independent. The main reason is that they are mostly implemented in GCS on the ground as shown in Fig. 2. That is, mostly the MDr flying for monitoring do the initial communications or necessary information processing whereas detailed analysis is done in the GCS due to power and hardware constraints. Hence, all the algorithms and technologies discussed here are equally valid for the four architectures discussed in this article. The only effect that will happen by varying the architecture is the increase in accuracy and reliability. For example, for localization under GCS-MA architecture, the technologies used for localization will be same but because of using known geographic positions of MDr network to estimate the position of ADr the results of localization would be more resilient to localization error and reliable. Similarly, this will happen for all other technologies discussed in this article.

*F. MDr Jamming & Hunting Technologies*

The ADr jamming can be done by using excess power and global positioning services (GPS) spoofing [14]. In literature, numerous jamming methods such as triangle method and genetic algorithm based jamming methods are proposed. Since, the MDr is moving, so these jamming methods are not too efficient. Hence, the cooperative jamming is hot topic and it can be more effective in MDr jamming because of its accuracy and efficiency. But these jamming schemes can generate the high interference to the MDr signal. So, in order reduce the effect of jamming on MDr, the power control, interference management algorithms, and beamforming algorithms are also desired to efficiently control the power of the jamming signals. After jamming the ADr, the hunting of the ADr can be done by using nets by taking care of security-sensitivity of the surrounding environment. For example, if the ADr is carrying some explosive material then the captured drone should be landed outside the security-sensitive area. Moreover, some efficient path-planning algorithms can be used to safely land the captured ADr.

## V. CHALLENGES AND FUTURE RESEARCH TRENDS IN MDR

*A. Interference Management Schemes for MDr*

For ADr monitoring, the MDr usually works on ISM bands that results in high co-channel interference with the existing wireless networks, which will significantly decrease the quality of service (QoS) for MDr. Also, the jamming technologies such as by using excess power and global positioning services (GPS) spoofing can generate the high interference to the MDr signal. So, Jamming signal power control algorithm needs to be designed to avoid surrounding MDr jamming. The existing schemes like Inter-cell interference coordination (ICIC) and coordinated multipoint transmission (CoMP) would not be sufficient to cope with it. The main reason is that the flying path of the MDr is not predictable as it is tracking the ADr whose flying path is not known in advance. In this situation, the power control schemes can work efficiently but need much modifications according to the architecture and environment. Hence, the future challenges include the development of novel and dynamically adaptive architecture-independent interference management schemes.

*B. Channel Models for MDr*

Channel varies as the environment and operational frequency changes. So, one of the important future research challenge is to design of channel models for MDr operational frequency band in order to test and validate different technologies under various architectures. The two main types of communications links in MDr are air-to-air and air-to-ground and that links should be reliable for safe operation of MDr. The reliability of these links can be achieved if we know the channel characteristics in advance. There is a major difference between the channel environment of air-to-ground and air-to-air communications in a sense of relaying. The MDr connected in FANET have to relay their information to the neighbor MDr in the network. Some theoretical models are proposed for air-to-air communications links but no properly tested and validated channel models exists for air-to-air communications links with relaying concepts. The existing channel models like two-ray and Rician channel model could be suitable for this environment. But experiments and tests are needed, to select the proper value of line-of-sight component factor $K$, to model the effect of Doppler effect for different MDr velocity as MDr has high Doppler shift due to high velocity, and for different Doppler spread. Moreover, more research is desired to test these channel models for different ranges of frequencies such as 433MHz, 1575.42MHz, and 2.4GHz. Furthermore, the more accurate 3D channel models design for MDr is an important future research challenges.

### C. Cooperative Schemes for MDr

Cooperation in MDr can be helpful to increase the reliability in detection, tracking, jamming, and mission and path planning. Our proposed GCS-SDN architecture will be suitable to implement cooperative schemes in MDr because of its ability to centrally controlled and managed technologies. By cooperation MDr can use path planning algorithms to produce a coordinated mission, which can increase the detection and tracking chances of ADr flying in their vicinity. Moreover, the cooperative MDr can be fault-tolerant, increase the detection and tracking range, cooperatively localize moving radio frequency source (e.g., ADr) with more accuracy [15], and reduce the interference by using coordinated scheduling. Hence, cooperative schemes is one of the important future research challenge.

## VI. CONCLUSION

The drones deployment is posing some serious security challenges to National heritage and commercial areas because they have potential to carry explosive materials. In this article, to combat with these security challenges we propose the concept of monitoring drones to detect, control, and jam the amateur drones. We also discuss the architecture for monitoring drones and categorize into three main types that is point-to-point architecture, cellular or satellite assisted architecture, and ad-hoc network architecture based on the security-sensitive situations. Moreover, some of the existing key technologies related to detection, tracking, routing, and jamming are discussed with their future research directions. Hence, the importance of deploying monitoring drones with some research and development challenges are highlighted in this article.

## REFERENCES

[1] Y. Zeng, R. Zhang, and T. J. Lim, "Wireless communications with unmanned aerial vehicles: opportunities and challenges," *IEEE Communications Magazine*, vol. 54, no. 5, pp. 36–42, 2016.

[2] M. Paul, S. M. Haque, and S. Chakraborty, "Human detection in surveillance videos and its applications-a review," *EURASIP Journal on Advances in Signal Processing*, vol. 2013, no. 1, p. 176, 2013.

[3] G. R. Rodríguez-Canosa, S. Thomas, J. del Cerro, A. Barrientos, and B. MacDonald, "A real-time method to detect and track moving objects (datmo) from unmanned aerial vehicles (uavs) using a single camera," *Remote Sensing*, vol. 4, no. 4, pp. 1090–1111, 2012.

[4] G. Mao, S. Drake, and B. D. Anderson, "Design of an extended kalman filter for uav localization," in *Information, Decision and Control. IDC'07*, 2007, pp. 224–229.

[5] X. Ge, S. Tu, G. Mao, C.-X. Wang, and T. Han, "5g ultra-dense cellular networks," *IEEE Wireless Communications*, vol. 23, no. 1, pp. 72–79, 2016.

[6] Z. Kaleem and K. Chang, "Public safety priority-based user association for load balancing and interference reduction in ps-lte systems," *IEEE Access*, vol. 4, pp. 9775–9785, 2016.

[7] A. Bradai, K. Singh, T. Ahmed, and T. Rasheed, "Cellular software defined networking: a framework," *IEEE Communications Magazine*, vol. 53, no. 6, pp. 36–43, 2015.

[8] "Aerospace forecasts 2016-2036," 2017. [Online]. Available: https://www.faa.gov

[9] B. Custers, *The Future of Drone Use: Opportunities and Threats From Ethical and Legal Perspectives.* Springer, 2016, vol. 27.

[10] Y. Saleem, M. H. Rehmani, and S. Zeadally, "Integration of cognitive radio technology with unmanned aerial vehicles: issues, opportunities, and future research challenges," *Journal of Network and Computer Applications*, vol. 50, pp. 15–31, 2015.

[11] L. Hauzenberger and E. Holmberg Ohlsson, "Drone detection using audio analysis," Master's thesis, 2015.

[12] A. Nemra and N. Aouf, "Robust ins/gps sensor fusion for uav localization using sdre nonlinear filtering," *IEEE Sensors Journal*, vol. 4, no. 10, pp. 789–798, 2010.

[13] O. K. Sahingoz, "Networking models in flying ad-hoc networks (fanets): Concepts and challenges," *Journal of Intelligent & Robotic Systems*, vol. 74, no. 1-2, p. 513, 2014.

[14] D. He, S. Chan, and M. Guizani, "Communication security of unmanned aerial vehicles," *IEEE Wireless Communications*, 2016.

[15] F. Koohifar, A. Kumbhar, and I. Guvenc, "Receding horizon multi-uav cooperative tracking of moving rf source," *IEEE Communications Letters*, vol. 21, no. 6, pp. 1433–1436, 2017.

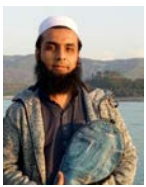

**Zeeshan Kaleem** Zeeshan Kaleem is Assistant Professor in the Department of Electrical Engineering, COMSATS Institute of information Technology, Pakistan. He received his Ph.D. degree in Electronics Engineering from INHA University, Korea (2016). He authored several peer-reviewed Journal/conference papers and holds 18 US/Korea patents. He is an Associate Editor in IEEE Access and served as Guest Editor in IEEE Communications Magazine. His research interest includes device-to-device (D2D) communications/discovery, unmanned air vehicles (UAV), resource allocation in fifth-generation (5G) networks.

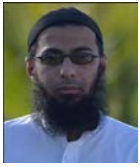 **Mubashir Husain Rehmani [M' 14-SM' 15]** Mubashir Husain Rehmani received his B.Eng. in computer systems engineering from MUET, Pakistan (2004) his M.S. from the University of Paris XI, France (2008), and his Ph.D. from the University Pierre and Marie Curie, Paris (2011). He is currently an assistant professor at COMSATS Institute of Information Technology, Pakistan. He received the certificate of appreciation, "Exemplary Editor of the IEEE Communications Surveys and Tutorials for the year 2015" from the IEEE Communications Society.

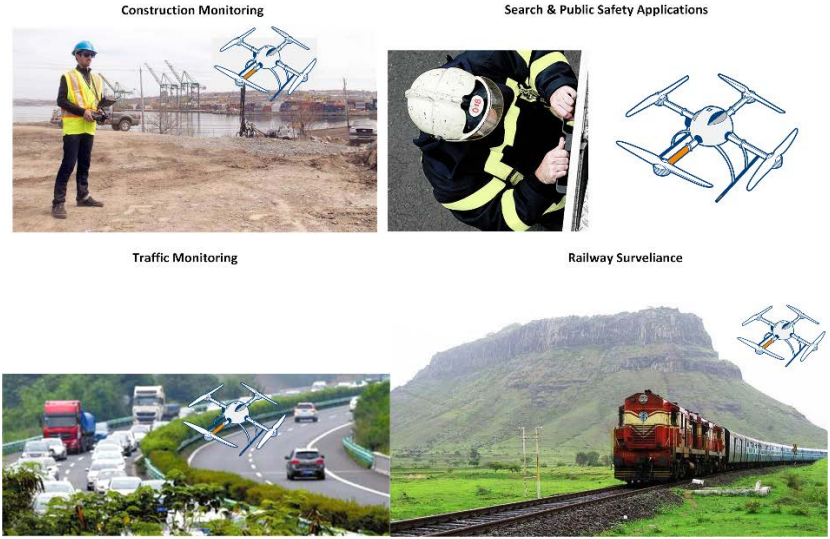

Fig. 1. Drones Applications: Construction monitoring, public-safety, traffic and railway monitoring.

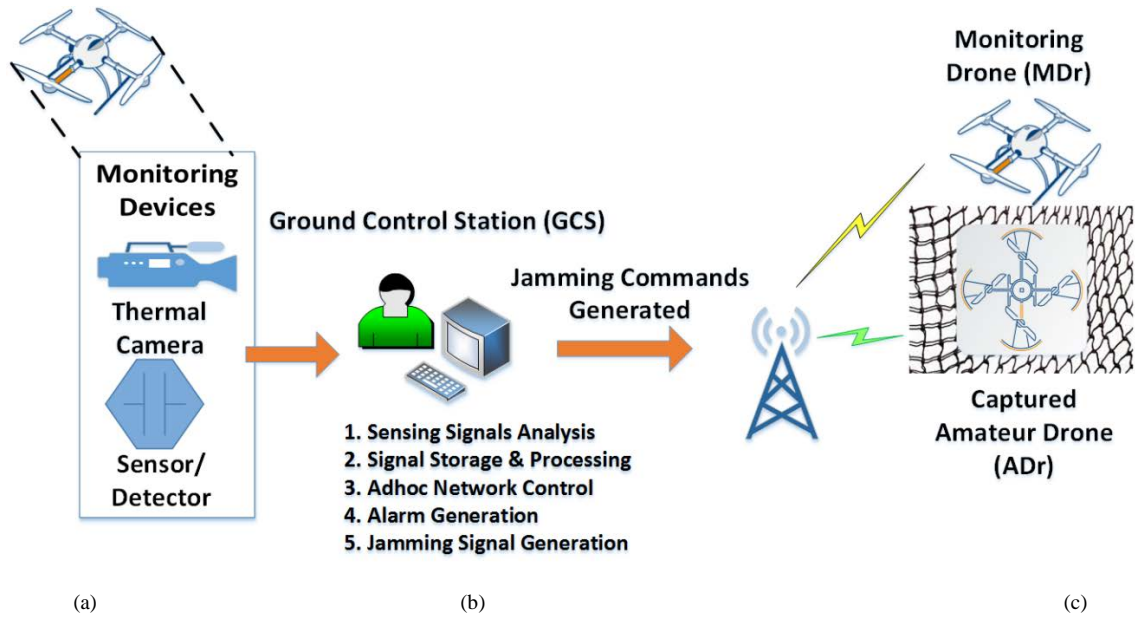

(a)     (b)     (c)

Fig. 2. Amateur drone capturing system step-wise representation: (a) data collection by using MDr; (b) ground control station for detection and jamming signal generation; (c) jamming and capturing of IDr.

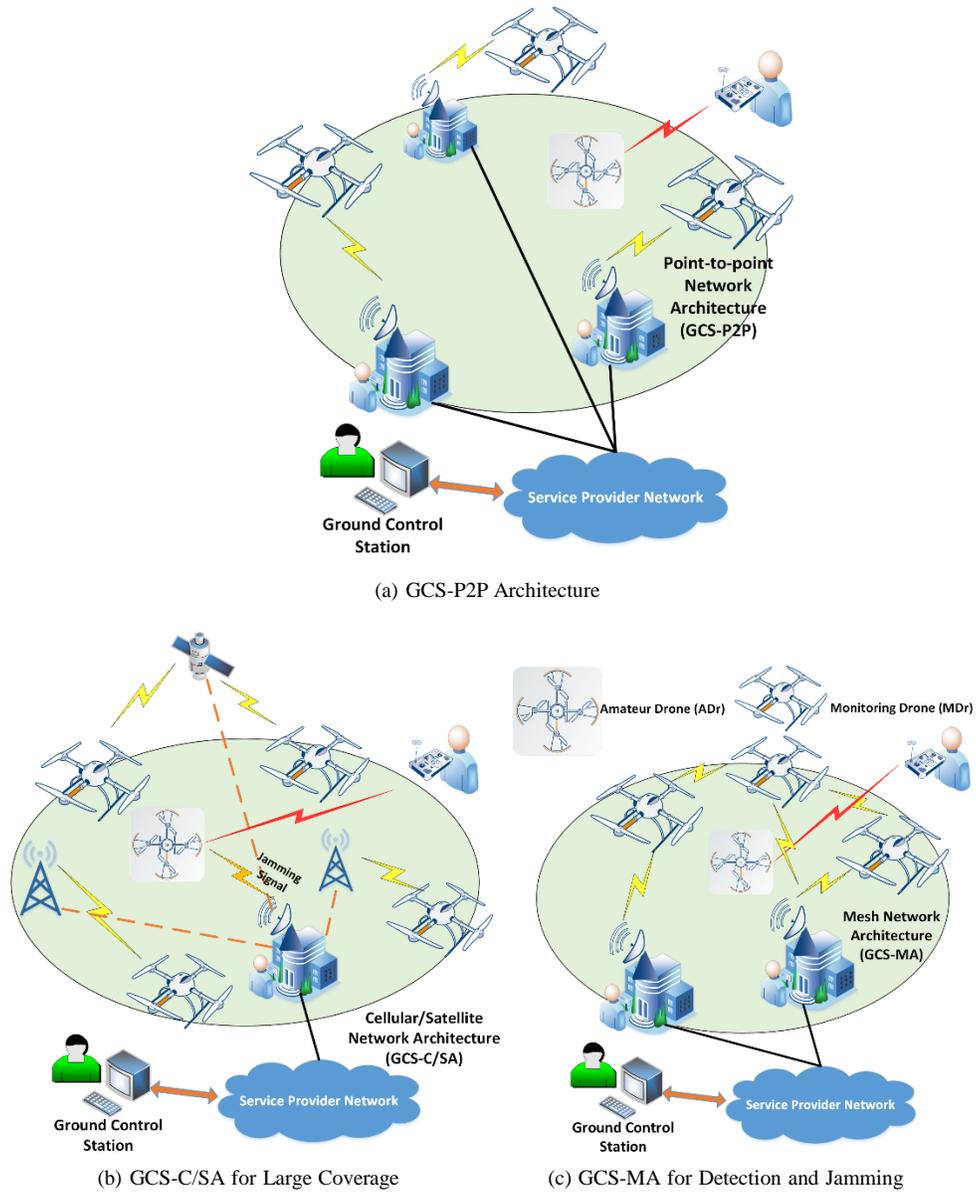

(a) GCS-P2P Architecture

(b) GCS-C/SA for Large Coverage    (c) GCS-MA for Detection and Jamming

Fig. 3. Monitoring drone architecture for controlling amateur drones flying in security-sensitive areas: (a) GCS-P2P architecture; (b) GCS-C/SA for large coverage; (c) GCS-MA for efficient detection and jamming.

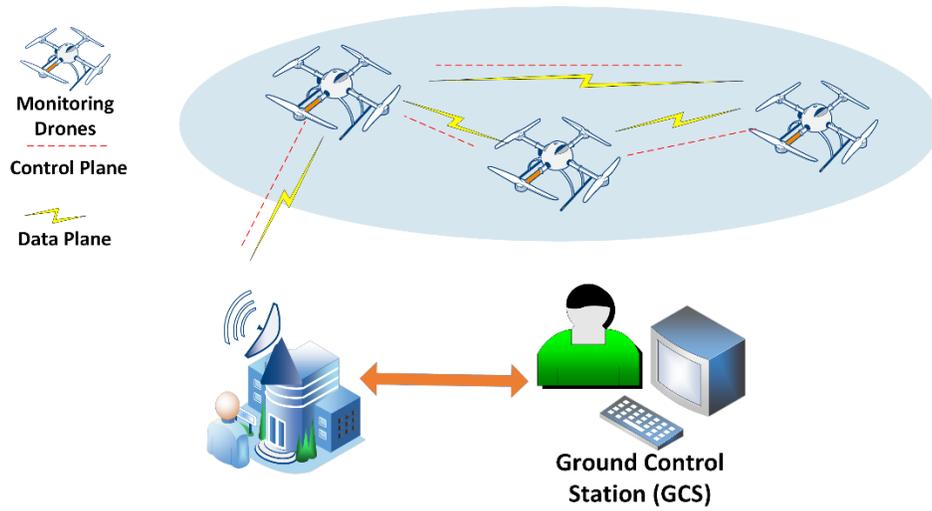

(a)   Conventional Deployment Scenario

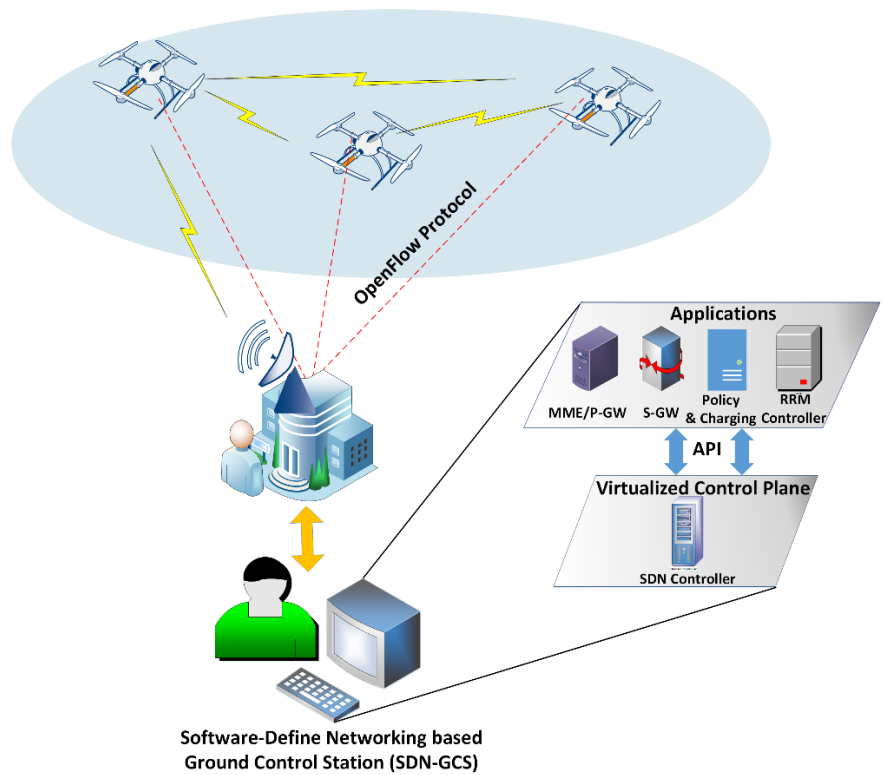

(b)   SDN Assisted MDr Deployment Scenario

Fig. 4.  MDr deployment scenarios: (a) Conventional deployment scenario; (b) SDN assisted MDr deployment scenario.

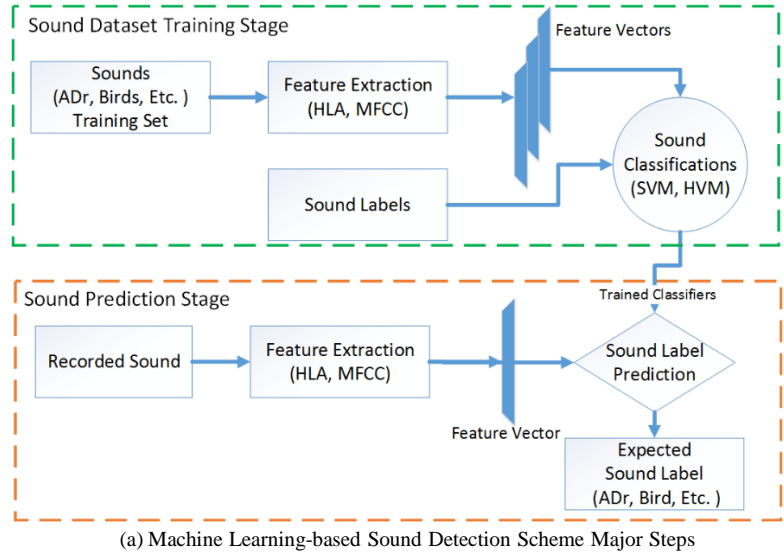

(a) Machine Learning-based Sound Detection Scheme Major Steps

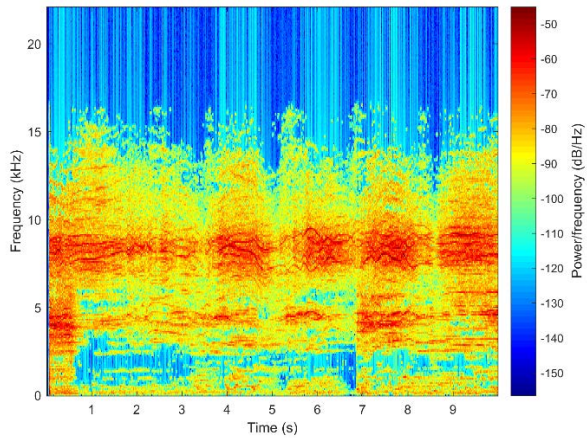

(b) Drone Spectrogram

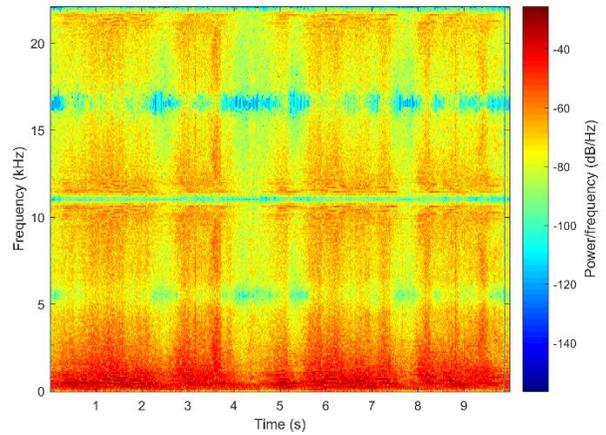

(c) Car Spectrogram

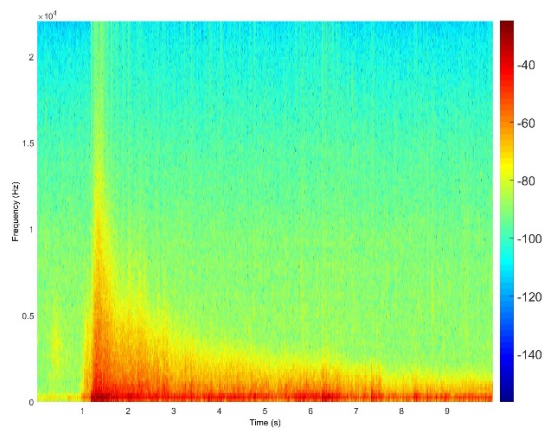

(d) Thunder Spectrogram

Fig. 5. Machine learning-based sound detection scheme and spectrograms for sound classifications: (a) Sound detection using machine learning; (b) drone spectrogram; (c) car spectrogram; (d) thunder spectrogram.

TABLE I
PROS AND CONS OF USING DIFFERENT FREQUENCY BANDS FOR MONITORING DRONE

| Frequency Band | Pros | Cons |
|---|---|---|
| 433MHz [9] | • Multiple channels available at one location for control and data<br>• Little multipath distortion<br>• Range is less susceptible to weather conditions<br>• Long transmission distance | • Does not support broadcast of quality video<br>• Legally transmitted power<br>inside band 433,050- 434,79 MHz<br>restricted to 10mW<br>effective radiated power (ERP) |
| 2.4GHz [9] | • Enables video broadcast<br>• Low cost system | • Expected interference in populated areas<br>due to the wide use: microwave<br>ovens, cordless phones, wireless LAN<br>• LOS operational requirements<br>• Range affected by humidity in the air<br>• Legal transmitting power restrictions<br>(100mW ERP) for the 802.11.b/g technology |
| 5.8GHz [9] | • Enables video broadcast<br>• Small transmitting antennas<br>• Low multipath distortions in spread spectrum OFDM modulation | • Severe multipath distortion causes<br>• Very poor performance in FM mode<br>• LOS operational requirements<br>• Range affected by humidity in the air |
| 1575.42MHz (GPS L1)<br>1227.60MHz (GPS L2)[9] | • Less interference with existing<br>Wireless communications system<br>• High coverage  • Available free of charge<br>without any subscription or license | • Ionosphere and troposphere leads to<br>slow down of signal propagation speed |
| Any Bands using<br>Cognitive Radios Technology [10] | •Dynamic spectrum access<br>• Suitable for public-safety<br>and emergency situations<br>• Opportunistic spectrum utilization based<br>on commercial applications requirements | • To handle the tradeoff between<br>sensing and transmission |